\documentclass[twocolumn,showpacs,preprintnumbers,amsmath,amssymb]{revtex4}
\usepackage{graphicx}
\usepackage{dcolumn}
\usepackage{bm}

\begin{document}

\title{
Spin-isospin response of deformed neutron-rich nuclei 
in a self-consistent Skyrme energy-density-functional approach}

\author{Kenichi Yoshida}
\affiliation{
Department of Physics, Graduate School of Science and Technology, 
Niigata University, Niigata 950-2181, Japan}

\date{\today}

\begin{abstract}
We develop a new framework of the self-consistent deformed proton-neutron 
quasiparticle-random-phase approximation (pnQRPA),  formulated 
in the Hartree-Fock-Bogoliubov (HFB) single-quasiparticle basis.
The same Skyrme force is used in both the HFB and pnQRPA calculations 
except in the proton-neutron particle-particle channel, 
where an $S=1$ contact force is employed. 
Numerical application is performed for  
Gamow-Teller (GT) strength distributions and $\beta$-decay rates 
in the deformed neutron-rich Zr isotopes located 
around the path of the rapid-neutron-capture process nucleosynthesis. 
It is found that 
the GT strength distributions are fragmented due to deformation. 
Furthermore we find that the momentum-dependent terms in the particle-hole residual 
interaction leads to a stronger collectivity of the GT giant resonance. 
The $T=0$ pairing enhances the low-lying strengths cooperatively with the $T=1$ pairing correlation, 
which shortens the $\beta$-decay half lives by at most an order of magnitude. 
The new calculation scheme reproduces well the observed isotopic dependence 
of the $\beta$-decay half lives of deformed $^{100-110}$Zr isotopes.
\end{abstract}



\maketitle

\section{Introduction}\label{sec1}

Study of unstable nuclei 
has been one of the major subjects in nuclear physics for a couple of decades. 
Collective mode of excitation emerged 
in the response of the nucleus to an external field 
is a manifestation of the interaction among nucleons. 
Thus, the spin-isospin channel of the interaction 
or the spin-isospin part of the energy-density functional (EDF), 
which is crusial 
for understanding and predicting the properties of 
unstable nuclei and asymmetric nuclear matter, 
has been much studied through especially the Gamow-Teller (GT) 
strength distributions~\cite{ost92,bak97}.

The GT strength distribution has been 
extensively investigated experimentally and theoretically 
not only because of interests in nuclear structure 
but also because $\beta$-decay half lives set a time scale 
of the rapid-neutron-capture process ($r$-process), 
and hence determine the production of heavy elements in the universe~\cite{lan03}. 
The $r$-process path is far away from the stability line, and 
involves neutron-rich nuclei. 
They are weakly bound and many of them are 
expected to be deformed according to 
the systematic Skyrme-EDF calculation~\cite{sto03}.

Collective modes of spin-isospin excitation in nuclei 
are described microscopically by 
the proton-neutron random-phase approximation (pnRPA) or 
the proton-neutron quasiparticle-RPA (pnQRPA) 
including the pairing correlations on top of 
the self-consistent Hartree-Fock (HF) or HF-Bogoliubov (HFB) mean fields 
employing the nuclear EDF. 
There have been many attempts to investigate the  spin-isospin 
modes of excitation in stable and unstable nuclei~\cite{paa07}. 
These studies are largely restricted to spherical systems, 
and the collective modes in deformed nuclei remain mostly unexplored.

The spin-isospin responses of deformed nuclei 
have been extensively investigated by 
the Madrid group~\cite{sar98, sar01a, sar12}
in connection to the studies of beta decay and double-beta decay 
in a Skyrme-pnQRPA model. 
The method employed in these preceding works 
relies on the BCS pairing instead of the HFB pairing, 
and the residual interactions
are treated in a separable approximation. 
The BCS approximation for pairing is inappropriate for 
describing the weakly bound nuclei due to the unphysical nucleon gas problem~\cite{dob84}. 
Furthermore, collectivity and details of the strength distribution 
are sensitive to both the shell structure around the Fermi levels 
and the residual interactions.
Quite recently, in Ref.~\cite{mut13}, 
the fully self-consistent Skyrme-pnQRPA model was established 
in an HFB single-canonical basis and was applied to the study of double-beta decay.

Recently, $\beta$-decay half lives of neutron-rich Kr to Tc isotopes with $A \simeq 110$ 
located on the boundary of the $r$-process path were newly measured 
at RIKEN RIBF~\cite{nis11}. 
The ground state properties such as deformation and superfluidity in 
neutron-rich Zr isotopes up to the drip line had been studied by employing 
the Skyrme-HFB method, and 
it had been predicted that Zr isotopes around $A=110$ 
are well deformed in the ground states~\cite{bla05}.

In the present article, 
to investigate the Gamow-Teller mode of excitation and 
$\beta$-decay properties in the deformed neutron-rich Zr isotopes, 
we construct a new framework of the calculation scheme 
employing the Skyrme EDF self-consistently 
in both the static and the dynamic levels. 
Furthermore, to describe properly the pairing correlations in weakly bound systems 
and coupling to the continuum states, 
the HFB equations are solved in the real space.
This framework is extended based on the 
deformed like-particle QRPA method developed in Ref.~\cite{yos08}.

The article is organized as follows: 
In Sec.~\ref{method}, the deformed Skyrme-HFB + pnQRPA method 
for describing the spin-isospin responses is explained. 
In Sec.~\ref{result}, results of the numerical analysis of the giant resonance 
in the neutron-rich Zr isotopes are presented. 
Discussion on effects of the $T=0$ pairing is included. 
Finally, summary is given in Sec.~\ref{summary}. 

\section{Theoretical Framework}\label{method}

\subsection{Microscopic calculation of spin-isospin modes of excitation in deformed nuclei}

To describe the nuclear deformation 
and the pairing correlations in the ground state, simultaneously, 
in good account of the continuum,
we solve the HFB equations~\cite{dob84,bul80}
\begin{align}
\begin{pmatrix}
h^{q}(\boldsymbol{r} \sigma)-\lambda^{q} & \tilde{h}^{q}(\boldsymbol{r} \sigma) \\
\tilde{h}^{q}(\boldsymbol{r} \sigma) & -(h^{q}(\boldsymbol{r} \sigma)-\lambda^{q})
\end{pmatrix}
\begin{pmatrix}
\varphi^{q}_{1,\alpha}(\boldsymbol{r} \sigma) \\
\varphi^{q}_{2,\alpha}(\boldsymbol{r} \sigma)
\end{pmatrix} \notag \\  
= E_{\alpha}
\begin{pmatrix}
\varphi^{q}_{1,\alpha}(\boldsymbol{r} \sigma) \\
\varphi^{q}_{2,\alpha}(\boldsymbol{r} \sigma)
\end{pmatrix} \label{HFB_eq}
\end{align}
in coordinate space using cylindrical coordinates $\boldsymbol{r}=(\rho,z,\phi)$.
We assume axial and reflection symmetries.
Here, the superscript $q$ denotes 
$\nu$ (neutron, $t_z = 1/2$) or $\pi$ (proton, $t_z =-1/2$).
The mean-field Hamiltonian $h$ is derived from the Skyrme EDF.
The pairing field $\tilde{h}$ is treated by using the density-dependent contact
interaction~\cite{cha76},
\begin{equation}
v_{\mathrm{pair}}(\boldsymbol{r}\sigma,\boldsymbol{r}^{\prime}\sigma^\prime)
=\dfrac{1-P_{\sigma}}{2}
\left[ t_{0}^{\prime}+\dfrac{t_{3}^{\prime}}{6}\varrho_{0}(\boldsymbol{r}) \right]
\delta(\boldsymbol{r}-\boldsymbol{r}^{\prime}), \label{pair_int}
\end{equation}
where $\varrho_{0}(\boldsymbol{r})$ denotes the isoscalar density
and $P_{\sigma}$ the spin exchange operator.

Since we consider the even-even mother (target) nuclei only, 
the time-reversal symmetry is assumed. 
A nucleon creation operator $\hat{\psi}^{\dagger}_q (\boldsymbol{r}\sigma)$ 
at the position $\boldsymbol{r}$ with the intrinsic spin $\sigma$ is then 
written in terms of the quasiparticle (qp) wave functions as
\begin{equation}
\hat{\psi}^{\dagger}_q (\boldsymbol{r}\sigma)
=\sum_{\alpha}\varphi_{1,\alpha}^q (\boldsymbol{r}\bar{\sigma})
\hat{a}^{\dagger}_{\alpha, q}
+\varphi_{2,\alpha}^{q *}(\boldsymbol{r}\sigma)\hat{a}_{\alpha, q}.
\end{equation}
The notation $\varphi(\boldsymbol{r}\bar{\sigma})$ is defined by 
$\varphi(\boldsymbol{r}\bar{\sigma})=-2\sigma \varphi(\boldsymbol{r} -\sigma)$. 

Using the quasiparticle basis obtained
as a self-consistent solution of the HFB equations (\ref{HFB_eq}),
we solve the pnQRPA equation
\begin{equation}
[\hat{H}^\prime, \hat{O}^\dagger_i] |0\rangle = \omega_i \hat{O}^\dagger_i |0\rangle,
\label{QRPA_eq}
\end{equation}
with $\hat{H}^\prime = \hat{H} -\lambda_\nu \hat{N}_\nu -\lambda_\pi \hat{N}_\pi$. 
The charge-changing QRPA phonon operators are defined as
\begin{equation}
\hat{O}^\dagger_{i} = \sum_{\alpha \beta}
X_{\alpha \beta}^i \hat{a}^\dagger_{\alpha,\nu}\hat{a}^\dagger_{\beta, \pi}
-Y_{\alpha \beta}^i \hat{a}_{\bar{\beta},\pi}\hat{a}_{\bar{\alpha},\nu},
\end{equation}
where $\hat{a}_{\bar{\alpha},q}$ is a quasiparticle annihilation operator of the 
time-reversed state of $\alpha$.

In the present calculation,
we solve the pnQRPA equation (\ref{QRPA_eq}) in the matrix
formulation
\begin{equation}
\sum_{\alpha^\prime \beta^\prime}
\begin{pmatrix}
A_{\alpha \beta \alpha^\prime \beta^\prime} & B_{\alpha \beta \alpha^\prime \beta^\prime} \\
B^*_{\alpha \beta \alpha^\prime \beta^\prime} & A^*_{\alpha \beta \alpha^\prime \beta^\prime}
\end{pmatrix}
\begin{pmatrix}
X_{\alpha^\prime \beta^\prime}^{i} \\ Y_{\alpha^\prime \beta^\prime}^{i}
\end{pmatrix}
=\omega_{i}
\begin{pmatrix}
1 & 0 \\ 0 & -1
\end{pmatrix}
\begin{pmatrix}
X_{\alpha \beta}^{i} \\ Y_{\alpha \beta}^{i}
\end{pmatrix} \label{eq:AB1}.
\end{equation}

Using the qp wave functions 
$\varphi_{1}(\boldsymbol{r}\sigma)$ and $\varphi_{2}(\boldsymbol{r}\sigma)$, 
the solutions of the coordinate-space HFB equation (\ref{HFB_eq}), 
the matrix elements of (\ref{eq:AB1}) are written as
\begin{align}
&A_{\alpha\beta\alpha^\prime\beta^\prime}=(E_{\alpha}+E_{\beta})\delta_{\alpha\alpha^\prime}\delta_{\beta\beta^\prime} 
+\int d1 d2 d1^\prime d2^\prime \notag \\ 
& \{ 
\varphi_{1,\alpha}^\nu(\boldsymbol{r}_{1}\bar{\sigma_{1}})
\varphi_{1,\beta}^\pi(\boldsymbol{r}_{2}\bar{\sigma_{2}})
\bar{v}_{\mathrm{pp}}(12;1^{\prime}2^{\prime})
\varphi^{\nu*}_{1,\alpha^\prime} (\boldsymbol{r}_{1}^{\prime}\bar{\sigma_{1}^{\prime}})
\varphi^{\pi*}_{1,\beta^\prime}(\boldsymbol{r}_{2}^{\prime}\bar{\sigma_{2}^{\prime}}) \notag \\
& +
\varphi_{2,\alpha}^\nu(\boldsymbol{r}_{1}\sigma_{1})
\varphi_{2,\beta}^\pi(\boldsymbol{r}_{2}\sigma_{2})
\bar{v}_{\mathrm{pp}}(12;1^{\prime}2^{\prime})
\varphi^{\nu*}_{2,\alpha^\prime} (\boldsymbol{r}_{1}^{\prime}\sigma_{1}^{\prime})
\varphi^{\pi*}_{2,\beta^\prime}(\boldsymbol{r}_{2}^{\prime}\sigma_{2}^{\prime}) \notag \\
&+
\varphi_{1,\alpha}^\nu(\boldsymbol{r}_{1}\bar{\sigma_{1}})
\varphi^{\pi*}_{2,\beta^\prime}(\boldsymbol{r}_{2}\sigma_{2})
\bar{v}_{\mathrm{ph}}(12;1^{\prime}2^{\prime})
\varphi_{2,\beta}^\pi(\boldsymbol{r}_{1}^{\prime}\sigma_{1}^{\prime})
\varphi^{\nu*}_{1,\alpha^\prime}(\boldsymbol{r}_{2}^{\prime}\bar{\sigma_{2}^{\prime}}) \notag \\
&+
\varphi_{1,\beta}^\pi(\boldsymbol{r}_{1}\bar{\sigma_{1}})
\varphi^{\nu*}_{2,\alpha^\prime}(\boldsymbol{r}_{2}\sigma_{2})
\bar{v}_{\mathrm{ph}}(12;1^{\prime}2^{\prime})
\varphi_{2,\alpha}^\nu(\boldsymbol{r}_{1}^{\prime}\sigma_{1}^{\prime})
\varphi^{\pi*}_{1,\beta^\prime}(\boldsymbol{r}_{2}^{\prime}\bar{\sigma_{2}^{\prime}})
\}, \label{A_matrix_QRPA} \\
&B_{\alpha\beta\alpha^\prime\beta^\prime}=
\int d1 d2 d1^\prime d2^\prime \notag \\ 
& \{-
\varphi_{1,\alpha}^\nu(\boldsymbol{r}_{1}{\bar \sigma_{1}})
\varphi_{1,\beta}^\pi(\boldsymbol{r}_{2}{\bar \sigma_{2}})
\bar{v}_{\mathrm{pp}}(12;1^{\prime}2^{\prime})
\varphi_{2,\bar{\alpha^\prime}}^\nu(\boldsymbol{r}_{1}^{\prime}\sigma_{1}^{\prime})
\varphi_{2,\bar{\beta^\prime}}^\pi(\boldsymbol{r}_{2}^{\prime}\sigma_{2}^{\prime}) \notag \\
&-
\varphi_{2,\alpha}^\nu(\boldsymbol{r}_{1}\sigma_{1})
\varphi_{2,\beta}^\pi(\boldsymbol{r}_{2}\sigma_{2})
\bar{v}_{\mathrm{pp}}(12;1^{\prime}2^{\prime})
\varphi_{1,\bar{\alpha^\prime}}^\nu(\boldsymbol{r}_{1}^{\prime}{\bar \sigma_{1}^{\prime}})
\varphi_{1,\bar{\beta^\prime}}^\pi(\boldsymbol{r}_{2}^{\prime}{\bar \sigma_{2}^{\prime}}) \notag \\
&-
\varphi_{1,\alpha}^\nu(\boldsymbol{r}_{1}{\bar \sigma_{1}})
\varphi_{1,\bar{\beta^\prime}}^\pi(\boldsymbol{r}_{2}{\bar \sigma_{2}})
\bar{v}_{\mathrm{ph}}(12;1^{\prime}2^{\prime})
\varphi_{2,\beta}^\pi(\boldsymbol{r}_{1}^{\prime}\sigma_{1}^{\prime})
\varphi_{2,\bar{\alpha^\prime}}^\nu(\boldsymbol{r}_{2}^{\prime}\sigma_{2}^{\prime}) \notag \\
&-
\varphi_{1,\beta}^\pi(\boldsymbol{r}_{1}{\bar \sigma_{1}})
\varphi_{1,\bar{\alpha^\prime}}^\nu(\boldsymbol{r}_{2}{\bar \sigma_{2}})
\bar{v}_{\mathrm{ph}}(12;1^{\prime}2^{\prime})
\varphi_{2,\alpha}^\nu(\boldsymbol{r}_{1}^{\prime}\sigma_{1}^{\prime})
\varphi_{2,\bar{\beta^\prime}}^\pi(\boldsymbol{r}_{2}^{\prime}\sigma_{2}^{\prime})
\}. \label{B_matrix_QRPA} 
\end{align}
Here, the time-reversed state is defined as
\begin{equation}
\varphi_{\bar{i}}(\boldsymbol{r}\sigma)=-2\sigma \varphi_{i}^{*}(\boldsymbol{r} -\sigma),
\end{equation}
and $\int d1$ stands for $\sum_{\sigma_1} \int d \boldsymbol{r}_1$.

If one assumes 
the effective interaction for the particle-hole (p-h) channel is local, 
$\bar{v}_{\mathrm{ph}}$ is written as
\begin{align}
&\bar{v}_{\mathrm{ph}}(12;1^{\prime}2^{\prime}) 
=V_{\mathrm{ph}}(\boldsymbol{r}_{1}\sigma_{1}\tau_1,\boldsymbol{r}_{2}\sigma_{2}\tau_2) \notag \\ 
& \times \delta(\boldsymbol{r}_{1}^{\prime}-\boldsymbol{r}_{1})
\delta_{\sigma_{1}^{\prime},\sigma_{1}}\delta_{\tau_1^{\prime},\tau_1}
\delta(\boldsymbol{r}_{2}^{\prime}-\boldsymbol{r}_{2})
\delta_{\sigma_{2}^{\prime},\sigma_{2}}
\delta_{\tau_2^{\prime},\tau_2},
\label{eq:v_ph}
\end{align}
and $V_{\mathrm{ph}}$ is
derived from the Skyrme EDF;
\begin{align}
&V_{\mathrm{ph}}
(\boldsymbol{r}_1\sigma_1 \tau_1,\boldsymbol{r}_2 \sigma_2 \tau_2) \notag \\ 
&= [v_{00}(\boldsymbol{r}_1)+ v_{01}(\boldsymbol{r}_1) \sigma_1 \cdot \sigma_2]
\tau_1 \cdot\tau_2 \delta(\boldsymbol{r}_1-\boldsymbol{r}_2) \notag \\
&+(v_{10}+v_{11}\sigma_1 \cdot \sigma_2) 
\tau_1 \cdot \tau_2 
 [\boldsymbol{k}^{\dagger 2}\delta(\boldsymbol{r}_1-\boldsymbol{r}_2)
+ \delta(\boldsymbol{r}_1-\boldsymbol{r}_2) \boldsymbol{k}^{2}] \notag \\
& + (v_{20}+v_{21}\sigma_1 \cdot\sigma_2 ) \tau_1 \cdot \tau_2 
 [\boldsymbol{k}^{\dagger}\cdot
 \delta(\boldsymbol{r}_1-\boldsymbol{r}_2)\boldsymbol{k} ] \notag \\
&+ v_{4} (\sigma_1 + \sigma_2)
\tau_1 \cdot\tau_2 \boldsymbol{k}^\dagger \times \boldsymbol{k},
\label{v_res_ph}
\end{align}
where $\boldsymbol{k}=(\overrightarrow{\nabla}_1-\overrightarrow{\nabla}_2)/2i$ and
$\boldsymbol{k}^\dagger=-(\overleftarrow{\nabla}_1-\overleftarrow{\nabla}_2)/2i$.
The coefficients in Eq.~(\ref{v_res_ph}) can be found 
in the appendix of Ref.~\cite{ter05} or in Ref~\cite{fra07}.

Assuming the proton-neutron particle-particle (p-p) effective interaction is local similarly 
to the p-h channel, 
we can write $\bar{v}_{\mathrm{pp}}$ as
\begin{align}
& \bar{v}_{\mathrm{pp}}(12;1^{\prime}2^{\prime})=
V_{\mathrm{pp}}(\boldsymbol{r}_{1}\sigma_{1} \tau_1,\boldsymbol{r}_{2}\sigma_{2} \tau_2) \notag \\ 
& \times \delta(\boldsymbol{r}_{1}^{\prime}-\boldsymbol{r}_{1})
\delta_{\sigma_{1}^{\prime},\sigma_{1}}\delta_{\tau_1^{\prime},\tau_1}
\delta(\boldsymbol{r}_{2}^{\prime}-\boldsymbol{r}_{2})
\delta_{\sigma_{2}^{\prime},\sigma_{2}} \delta_{\tau_2^{\prime},\tau_2}.
\end{align}
The residual interaction in the p-p channel $V_{\mathrm{pp}}$ 
could be derived from the proton-neutron pairing EDF. 
However, it is not well established yet. 
What we need in our framework is an interaction between the 
proton-neutron particle-particle (p-p), and hole-hole (h-h) pairs.
In the present calculation, we consider the p-p (h-h) interaction between 
the $T=0, S=1$ pair only
\begin{align}
V_{\mathrm{pp}}(\boldsymbol{r}_1 \sigma_1 \tau_1, 
\boldsymbol{r}_2 \sigma_2 \tau_2)
=
\dfrac{1+P_\sigma}{2} \dfrac{1-P_\tau}{2}
v(\boldsymbol{r}_1) \delta(\boldsymbol{r}_1-\boldsymbol{r}_2)
\end{align} 
and take $v(\boldsymbol{r})=v_0$ as a constant for simplicity. 
Here, $P_\tau$ denotes the isospin exchange operator.

The GT$^{\pm}$ transition strengths to the state $i$ with angular momentum $K (K=0, \pm1)$ 
are calculated as 
\begin{align}
B(\mathrm{GT}^\pm; i) &= \dfrac{g_A^2}{4\pi} | \langle i | \hat{F}_K^\pm |0 \rangle  |^2, 
\label{B_GT}
\\ 
\langle i | \hat{F}_K^\pm |0 \rangle
&=\sum_{\alpha \beta} 
X_{\alpha \beta}^i \langle \alpha \beta| \hat{F}_K^\pm| \mathrm{HFB}\rangle
-Y_{\alpha \beta}^i  \langle \alpha \beta| \hat{F}_K^\mp | \mathrm{HFB}\rangle
\end{align}
under the quasi-boson approximation.
The HFB vacuum is denoted as $|\mathrm{HFB}\rangle$, 
and 
$|\alpha \beta \rangle 
= \hat{a}^\dagger_{\alpha,\nu} \hat{a}^\dagger_{\beta,\pi} |\mathrm{HFB}\rangle$ 
is a 2qp excited state.
The GT$^\pm$ operators are given by
\begin{align}
\hat{F}_K^+ &=\sum_{\sigma \sigma^\prime} \int d\boldsymbol{r}
\hat{\psi}^\dagger_\nu (\boldsymbol{r}\sigma^\prime) 
\langle \sigma^\prime|\sigma_K|\sigma\rangle 
\hat{\psi}_\pi(\boldsymbol{r}\sigma),
\\
\hat{F}_K^- &= \sum_{\sigma \sigma^\prime} \int d\boldsymbol{r}
\hat{\psi}^\dagger_\pi (\boldsymbol{r}\sigma^\prime) 
\langle \sigma^\prime|\sigma_K|\sigma\rangle 
\hat{\psi}_\nu(\boldsymbol{r}\sigma). \label{eq:GT-}
\end{align}

The transition-strength distributions as functions of the 
excitation energy $E^*$ with respect to 
ground state of 
the odd-odd daughter nucleus 
are calculated as
\begin{equation}
\label{S_l}
R^\pm (E^*)
=\sum_{K}\sum_{i} \dfrac{\gamma/2}{\pi}\dfrac{ | \langle i | \hat{F}_K^\pm |0 \rangle  |^2 }
{[E^*- (\omega_{i}-E_0)]^{2}+\gamma^{2}/4}.
\end{equation} 
The smearing width $\gamma$ is introduced to make the strength distributions easier to read. 
$E_0$ denotes the lowest quasiparticle energy of protons and neutrons. 
When either or both pairing gaps vanish, 
we take the lowest occupied neutron and unoccupied proton states 
for the $t_-$ channel.
It is noted that the spin-parity of the state with $E_0$ is, in general, 
different from $1^+$.

\subsection{Details of the numerical calculation}
We employ the SkM*~\cite{bar82} and SLy4~\cite{cha98} EDFs for 
the mean-filed Hamiltonian and the 
residual interaction for the p-h channel.
The pairing strength parameter $t_{0}^{\prime}$ is
determined so as to approximately reproduce the experimental pairing gap of
$^{120}$Sn ($\Delta_{\mathrm{exp}}=1.245$ MeV) as in Ref.~\cite{yos10}, 
where the giant monopole resonance in the deformed neuron-rich Zr isotopes was investigated.
The strengths $t_{0}^{\prime}=-240$ and $-290$ MeV fm$^{3}$ for the
mixed-type interaction ($t_{3}^{\prime}=-18.75t_{0}^{\prime}$)~\cite{ben05} 
lead to the neutron pairing gap
$\langle \Delta_{\nu}\rangle=1.20$ and 1.24 MeV in $^{120}$Sn 
with the SkM* and SLy4 EDFs, respectively.
The strength parameter $v_0$ for the $T=0$ pairing interaction 
can be considered as a free parameter, 
because it dose not affect the ground state properties, and it is active only in the 
dynamic level. Our procedure to determine it is to fit 
approximately the $\beta$-decay lifetime of 
$^{100}$Zr $(T_{1/2}^{\mathrm{exp}}=7.1$ s~\cite{Zr100}). 
The strengths $v_0=-395$ and $-320$ MeV fm$^3$ give 
the calculated $\beta$-decay half life $T_{1/2} = 7.08$ s and 7.63 s 
with the SkM* and SLy4 EDFs, respectively.

Because of the assumption of the axially symmetric potential,
the $z-$component of the qp angular momentum, $\Omega$,
is a good quantum number. 
Assuming time-reversal symmetry and reflection symmetry with respect to the $x-y$ plane,
we have only to solve Eq.~(\ref{HFB_eq})
for positive $\Omega$ and positive $z$.
We use the lattice mesh size $\Delta\rho=\Delta z=0.6$ fm and a box
boundary condition at $\rho_{\mathrm{max}}=14.7$ fm, $z_{\mathrm{max}}=14.4$ fm 
to discretize the continuum states.
The differential operators are represented by use 
of the 13-point formula of finite difference method. 
The quasiparticle energy cutoff is chosen at $E_{\mathrm{qp,cut}}=60$ MeV
and the quasiparticle states up to $\Omega^{\pi}=31/2^{\pm}$ are included.

We introduce an additional truncation for the pnQRPA calculation,
in terms of the 2qp energy as
$E_{\alpha}+E_{\beta} \leq 60$ MeV.
This reduces the number of 2qp states to, for instance,
about 30 000 for the $K^\pi=0^+$ excitation.
The number of 2qp states included  in the calculation 
is large enough to satisfy the Ikeda sum-rule values to an accuracy of 1\%. 
The calculation of the QRPA matrix elements in the qp basis, 
and diagonalization of the QRPA matrix 
are performed in the parallel computers as in Ref.~\cite{yos13}. 

As a test calculation, our method is applied to the isobaric analogue state (IAS) 
in $^{90}$Zr. When the Coulomb potential is discarded, 
that is, the electron charge $e$ is set zero, 
the IAS appears at $\sim$ 0.2 MeV excitation energy with respect to the 
ground state of $^{90}$Zr with the SkM* EDF and 
the 2qp energy cutoff described above. 
For the deformed $^{110}$Zr case, we obtained 
the IAS at $\sim$ 0.4 MeV. 
This implies that our calculation scheme satisfies the self-consistency 
between the static and the dynamic calculations~\cite{aue80}. 
It is noted here that the mean energy of the IAS is given by
\begin{equation}
\langle E_{\mathrm{IAS}}\rangle 
= e^2 \int d\boldsymbol{r} \int d \boldsymbol{r}^\prime
\dfrac{\varrho_\pi(\boldsymbol{r})[ \varrho_\nu(\boldsymbol{r}^\prime) 
-\varrho_\pi (\boldsymbol{r}^\prime) ] }
{(N-Z)|\boldsymbol{r}-\boldsymbol{r}^\prime |}
\end{equation}
with the sum-rule method for the separable interaction~\cite{ost92}.
Thus, without the Coulomb potential, 
the excitation energy would be zero 
if and only if the isospin is a good quantum number~\cite{aue80}. 
The finite excitation energy obtained here is due to the spurious 
isospin mixing in the HFB approximation for $N \ne Z$ nuclei.

\begin{table*}[t]
\begin{center}
\caption{Ground state properties of Zr isotopes 
obtained by the deformed HFB calculation 
with the SkM* and the mixed-type pairing ($t_0^\prime=-240$ MeV fm$^3$) interactions. 
Chemical potentials $\lambda_q$, deformation parameters $\beta_2^q$,
average pairing gaps $\langle \Delta \rangle_q$,
root-mean-square radii $\sqrt{\langle r^{2} \rangle_q}$ for 
neutrons and protons, the lowest 2qp excitation energy $E_0$, 
and $Q$ values of $\beta$ decay are listed.}
\label{GS_SkM}
\begin{tabular}{crrrrrrrr}
\hline \hline
\noalign{\smallskip}
 & $^{98}$Zr  & $^{100}$Zr & $^{102}$Zr & $^{104}$Zr 
 & $^{106}$Zr & $^{108}$Zr & $^{110}$Zr  & $^{112}$Zr  \\
\noalign{\smallskip}\hline\noalign{\smallskip}
$\lambda_{\nu}$ (MeV)  & $-5.68$  & $-6.91$ & $-6.14$  &  $-5.51$
& $-4.88$ & $-4.57$ & $-4.31$  & $-3.76$  \\
$\lambda_{\pi}$ (MeV)  & $-10.0$ & $-10.6$ & $-11.4$ & $-12.2$
& $-12.9$ & $-13.6$ & $-14.3$  & $-15.0$ \\
$\beta_{2}^{\nu}$  & 0.00  & 0.38 & 0.39 & 0.39
& 0.38 & 0.37 & 0.37 & 0.39   \\
$\beta_{2}^{\pi}$  & 0.00 & 0.41 & 0.43 & 0.43
& 0.42 & 0.42 & 0.42 & 0.43  \\
$\langle \Delta \rangle_{\nu}$ (MeV) & 0.63  & 0.00 & 0.51 & 0.00
& 0.54 & 0.78 & 0.55 & 0.00  \\
$\langle \Delta \rangle_{\pi}$ (MeV) & 0.00 & 0.32 & 0.00 & 0.00
& 0.00 & 0.00 & 0.00 & 0.00   \\
$\sqrt{\langle r^{2} \rangle_{\nu}}$ (fm)  & 4.47  & 4.60 & 4.65 & 4.70
& 4.74 & 4.77 & 4.82 & 4.88  \\
$\sqrt{\langle r^{2} \rangle_{\pi}}$ (fm)  & 4.28  & 4.43 & 4.46 & 4.48
& 4.50 & 4.51 & 4.53 & 4.56 \\
$E_0$ (MeV) & 1.93 & 1.20 & 1.09 & 1.48 & 1.32 & 1.45 & 1.47 & 1.89 \\   
$Q_\beta$ (MeV) & 3.17 & 3.27 & 4.95 & 5.99 & 7.48 & 8.36 & 9.30 & 10.1 \\
\noalign{\smallskip}
\hline \hline
\end{tabular}
\end{center}
\end{table*}

\begin{table*}[t]
\begin{center}
\caption{Same as Table~\ref{GS_SkM} but obtained with the SLy4 and 
the mixed-type pairing ($t_0^\prime=-290$ MeV fm$^3$) interactions.}
\label{GS_SLy4}
\begin{tabular}{crrrrrrrr}
\hline \hline
\noalign{\smallskip}
 & $^{98}$Zr  & $^{100}$Zr & $^{102}$Zr & $^{104}$Zr 
 & $^{106}$Zr & $^{108}$Zr & $^{110}$Zr  & $^{112}$Zr  \\
\noalign{\smallskip}\hline\noalign{\smallskip}
$\lambda_{\nu}$ (MeV)  & $-5.52$  & $-6.05$ & $-5.41$  &  $-4.90$
& $-4.48$ & $-4.00$ & $-3.53$  & $-2.90$  \\
$\lambda_{\pi}$ (MeV)  & $-11.1$ & $-11.4$ & $-12.2$ & $-13.0$
& $-13.9$ & $-14.8$ & $-15.5$  & $-16.4$ \\
$\beta_{2}^{\nu}$  & 0.00  & 0.38 & 0.38 & 0.39
& 0.38 & 0.36 & 0.38 & 0.40   \\
$\beta_{2}^{\pi}$  & 0.00 & 0.40 & 0.41 & 0.43
& 0.42 & 0.41 & 0.42 & 0.44  \\
$\langle \Delta \rangle_{\nu}$ (MeV) & 1.04  & 0.60 & 0.70 & 0.71
& 0.72 & 0.54 & 0.65 & 0.00  \\
$\langle \Delta \rangle_{\pi}$ (MeV) & 0.00 & 0.62 & 0.53 & 0.27
& 0.00 & 0.33 & 0.00 & 0.00   \\
$\sqrt{\langle r^{2} \rangle_{\nu}}$ (fm)  & 4.48  & 4.62 & 4.66 & 4.70
& 4.74 & 4.77 & 4.83 & 4.89  \\
$\sqrt{\langle r^{2} \rangle_{\pi}}$ (fm)  & 4.29  & 4.44 & 4.47 & 4.49
& 4.51 & 4.53 & 4.56 & 4.59 \\
$E_0$ (MeV) &  2.69 & 1.48 & 1.41 & 1.42 & 1.59 & 1.69 & 1.83 & 1.74 \\  
$Q_\beta$ (MeV) & 3.67 & 4.65 & 6.16 & 7.46 & 8.61 & 9.89 & 10.9 & 12.5 \\ 
\noalign{\smallskip}
\hline \hline
\end{tabular}
\end{center}
\end{table*}

As stated in Introduction, 
a similar calculation of the self-consistent HFB + pnQRPA 
for axially deformed nuclei has been recently reported~\cite{mut13}.
They adopt the canonical-basis representation 
and introduce a further truncation according to the occupation probabilities 
of 2qp excitations. 
In contrast, we adopt the qp representation 
and truncation simply due to the 2qp energies. 

\section{Results and Discussion}\label{result}
\subsection{Ground state properties}\label{result_gs}

In Tables~\ref{GS_SkM} and \ref{GS_SLy4}, we summarize the ground state 
properties of the Zr isotopes 
calculated with the SkM* and SLy4 EDFs combined with the mixed-type 
pairing interaction. 
The ground state of $^{98}$Zr is spherical, and we see a sudden onset of 
deformation in the Zr isotopes with $N \ge 60$.
Both Skyrme EDFs give the similar deformations and root-mean-square radii 
of the ground states in nuclei under investigation. 
In the present article, we investigate the GT excitation mainly in deformed nuclei.
However, to see the deformation effect, we include 
$^{98}$Zr as a reference.

The pairing properties 
calculated with the two EDFs seem rather different; 
the SkM* combined with the mixed-type pairing interaction 
gives weaker pairing correlations. 
For instance, the pairing gap of neutrons in $^{100}$Zr vanishes, 
and both neutrons and protons are unpaired in $^{104}$Zr with SkM*.  
For $^{112}$Zr, both neutrons and protons are calculated to be unpaired 
consistently with the SkM* and SLy4 EDFs. 
The deformed shell gap of neutrons formed by between [402]5/2 and [541]1/2 orbitals 
and of protons formed by between [431]1/2 and [422]5/2 orbitals are 
1.39 (2.01) MeV and 1.96 (1.64) MeV, respectively with SkM* (SLy4).

For each isotopes in the tables, the neutron chemical potential is shallower, 
and the proton chemical potential is instead deeper with SLy4 than with SkM*. 
This gives larger $Q_\beta$ values of $\beta$-decay with the SLy4 EDF 
because the differences in $E_0$ are not very large. 
The $Q_\beta$ value is given by
\begin{align}
Q_\beta &= M(A,Z) - M(A,Z+1) -m_e \notag \\
&= (m_n-m_p-m_e)+B(A,Z+1)-B(A,Z) \notag \\
& \simeq \Delta M_{n-H} + \lambda_\nu - \lambda_\pi - E_0
 \end{align}
under the independent quasi-particle approximation. 
 Here $\Delta M_{n-H}=0.78227$ MeV is the mass difference between 
a neutron and a hydrogen atom.  
Noted that $Q_\beta$ is given by 
$\Delta M_{n-H} - \min(\varepsilon_\pi^{\mathrm{unocc.}} - \varepsilon_\nu^{\mathrm{occ.}})$ 
in terms of the single-particle energies for unpaired systems if we choose $E_0$ as described above.

\subsection{GT giant resonance}

\begin{figure}
\begin{center}
\includegraphics[scale=0.48]{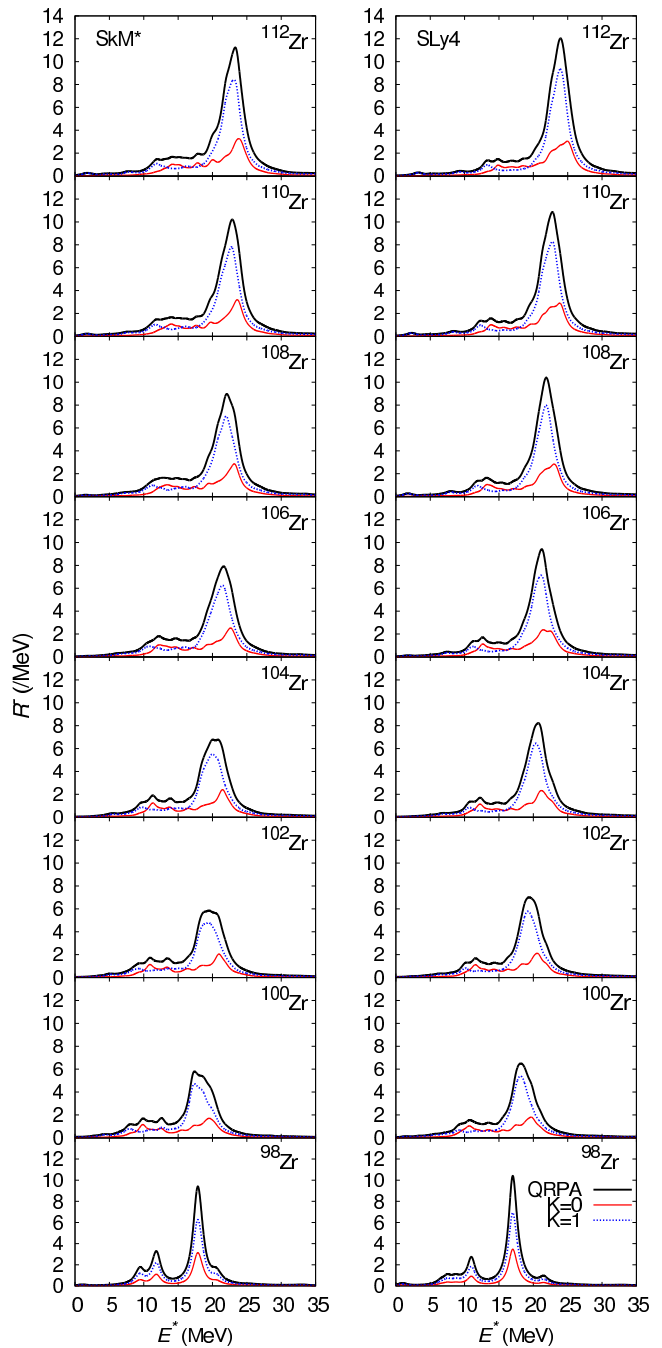}
\caption{Strength distributions $R^- (E^*)$ as functions of the excitation energy of daughter nuclei. 
The SkM* (left) and SLy4 (right) EDFs combined with  the mixed-type pairing interaction 
are employed for the calculation. The smearing width $\gamma$ is set $1$ MeV.
\label{fig:GT}} 
\end{center}
\end{figure}

Figure~\ref{fig:GT} shows the strength distributions associated with the 
GT$^-$ operator (\ref{eq:GT-}) without the $T=0$ pairing 
in the Zr isotopes as functions of the excitation energy with respect to the daughter nuclei. 
We also show in this figure the contribution of the $K=0$ and $1$ components to the total strength. 
As we will discuss in Sec.~\ref{T=0_pairing}, 
we see a tiny amount of energy change 
due to $T=0$ pairing for the GT giant resonance (GR). 
We need to multiply the strengths shown in the figure 
by $g_A^2/4\pi$ or 
$(g_A^2)_\mathrm{eff}/4\pi$ including the quenching effect 
to obtain the $B(\mathrm{GT}^-)$ values in Eq.~(\ref{B_GT}).

\begin{figure}
\begin{center}
\includegraphics[scale=0.32]{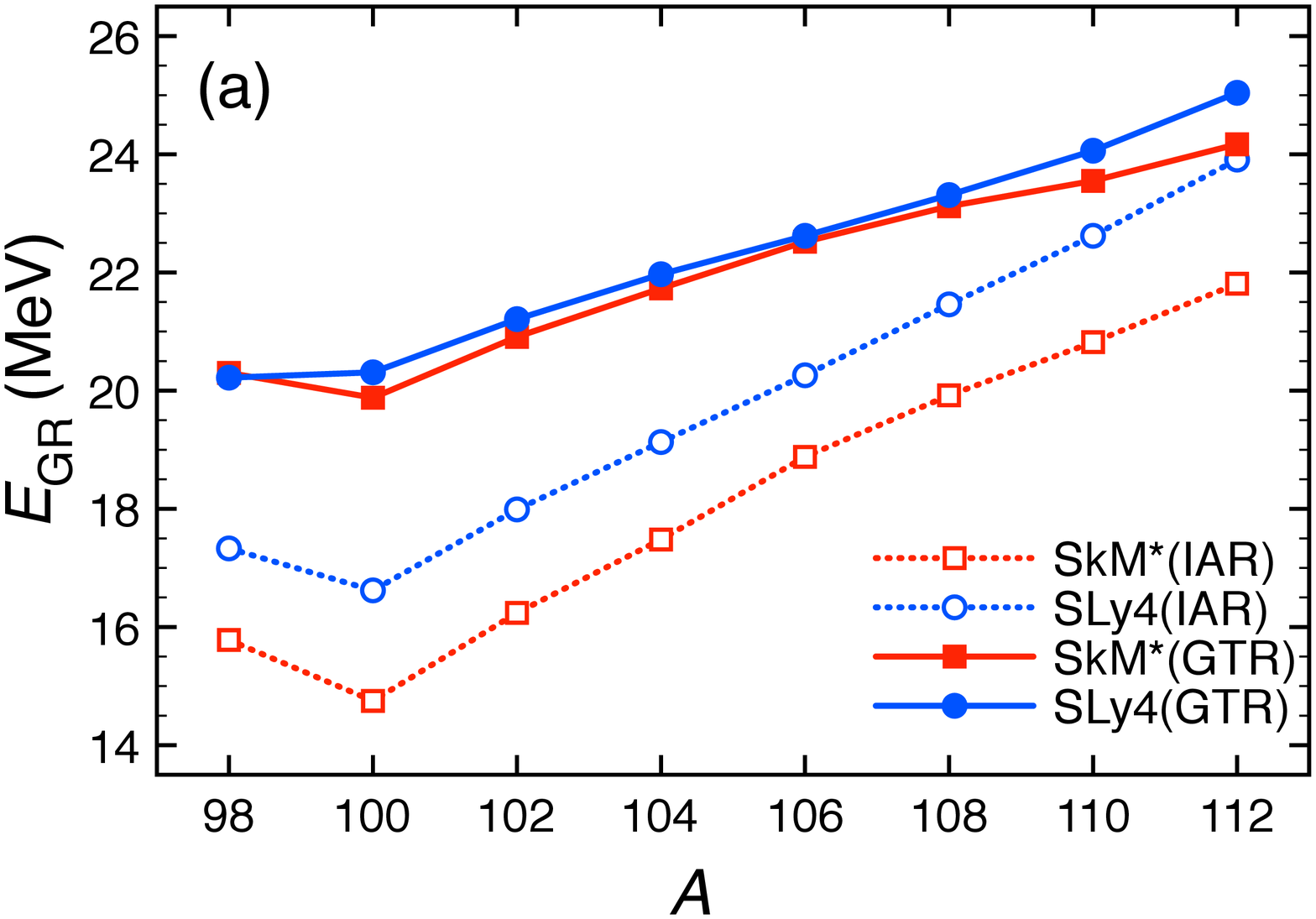}
\includegraphics[scale=0.32]{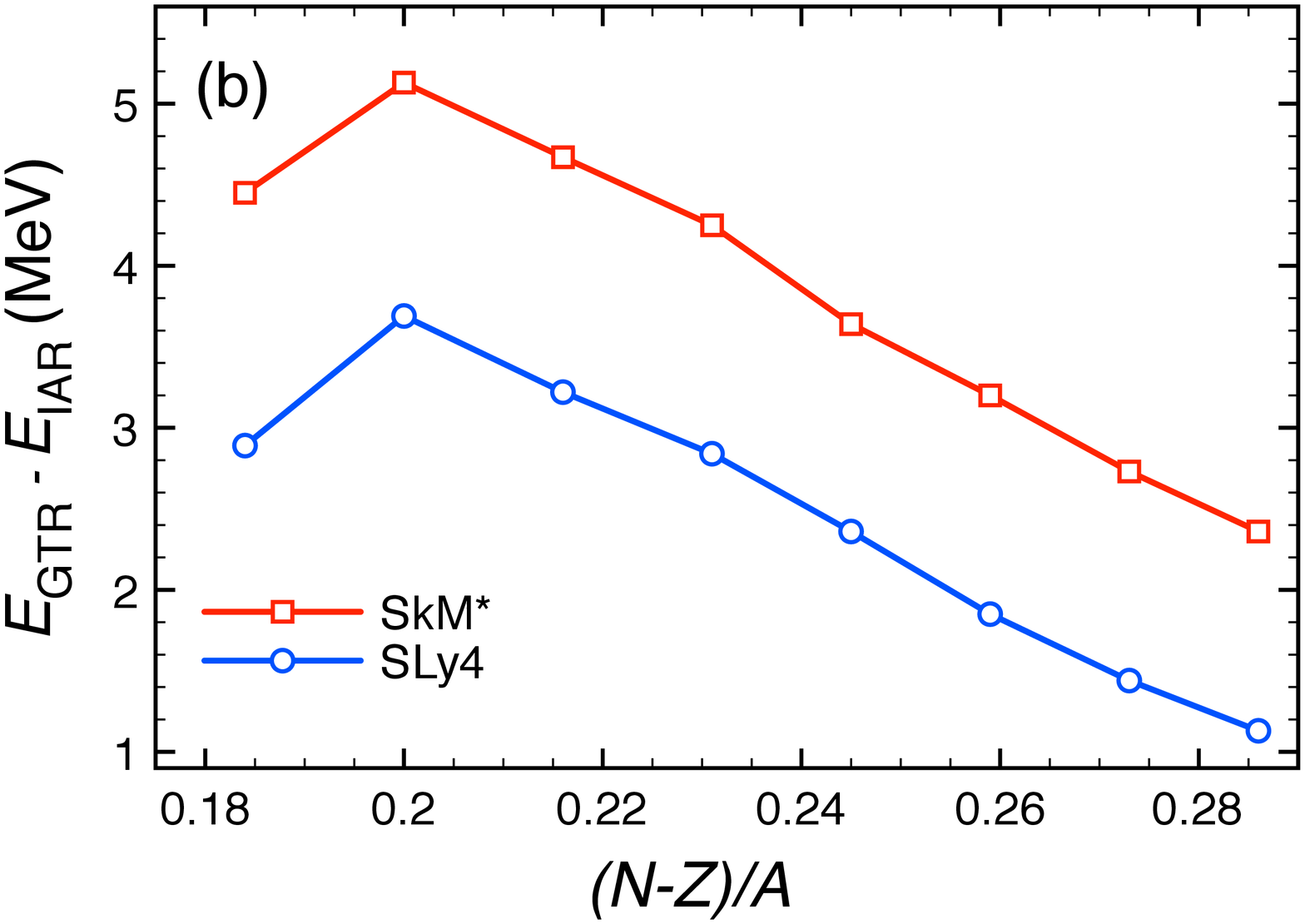}
\caption{
(a): Centroid energies of the Gamow-Teller (GT) and the isobaric analogue (IA) 
resonances of the Zr isotopes 
as functions of the mass number calculated with the SkM* and 
SLy4 EDFs. 
(b): Energy differences of the GTR and IAR 
for the Zr isotopes as functions of the relative neutron excess.
\label{fig:GR_energy}} 
\end{center}
\end{figure}

To quantify the excitation energy of the GR,
we introduce the centroid energy 
which is frequently used in the experimental analysis,
defined by
\begin{equation}
E_c=\dfrac{m_1}{m_0},
\end{equation}
where $m_k$ is a $k-$th moment of 
the transition-strength distribution in an energy interval of $[E_a, E_b]$ MeV;
\begin{equation}
\label{m_k_con}
m_k  \equiv  
\sum_K \sum_{E_a < E_i^* < E_b} E_i^{* k}
| \langle i | \hat{F}_K^\pm |0 \rangle  |^2,
\end{equation} 
with $E_i^* = \omega_i -E_0$.

In Fig.~\ref{fig:GR_energy}(a), we show the centroid energies of the GTGR and 
the isobaric analogue resonance (IAR). 
Since we have only a single peak of the IAR, 
we take the energy interval to evaluate the centroid energy as 
$E_a=0$ and $E_b=40$ MeV. 
For evaluation of the centroid energy of the GTGR, 
we take $E_a=15$ and $E_b=30$ MeV.

The centroid energies obtained 
with the SkM* and SLy4 EDFs
of the GTGR are similar to each other. 
Properties of a GR are liked to the nuclear matter properties. 
In nuclear matter, the interaction strength in the GT channel is sensitive to 
the Landau-Migdal (LM) parameter $g_0^\prime$~\cite{gia81}.
The SkM* and SLy4 EDFs give the similar values of $g_0^\prime$; 0.94 and 0.90, respectively. 
Thus, the collectivity of the GTGR calculated may be similar. 
These values of $g_0^\prime$ 
are large and comparable with $g_0^\prime = 0.93$ 
of the SGII interaction~\cite{gia81,eng99}, 
which is designed to describe the spin-isospin excitations.
Noted that, however, they are much smaller than the empirical value 
$g_0^\prime(\mathrm{exp}) \simeq 1.8$~\cite{ost92}.

We are going to discuss the deformation effects. 
In the spherical $^{98}$Zr nucleus, 
the centroid energy of the GTGR is about 20 MeV, 
and it appears as a narrow peak. 
Besides the GTGR, 
we see a low-energy resonance structure in the energy region of $5-15$ MeV. 
When the system gets deformed, strengths both of 
the GTGR and of the low-energy resonance are fragmented. 
The deformation splitting between the $K=0$ and 1 states 
of the GTGR is at most 1 MeV. 
So, the splitting effect is washed out by the smearing width $\gamma$. 
This is consistent with the finding in Ref.~\cite{urk01}, 
where the schematic residual interaction was employed. 
The spreading effect $\Gamma^{\downarrow}$ not taken in the present calculation 
may be larger than $1$ MeV so that it is difficult to observe the deformation splitting 
of the GTGR experimentally. 

As increasing the neutron number, the centroid energy of the GTGR 
monotonically increases. 
This characteristic feature of increase in the excitation energy 
as a function of the neutron (mass) number 
is also found in Ref.~\cite{sar10}.  
In $^{112}$Zr, the centroid energy reaches about 25 MeV.
We find also a monotonic increase in the centroid energy of the IAR 
in the deformed systems.
Seen is a general feature that 
the energies of the IAR calculated with SLy4 are higher than 
those calculated with SkM*. 
It is noted that there has been a discussion on the 
correlation between the symmetry energy and the energy of the IAR~\cite{dan13}. 
Indeed, the symmetry coefficient of SLy4 (32.00 MeV) 
is larger than that of SkM* (30.03 MeV).

With the sum-rule method for the separable interaction~\cite{ost92}, 
the energy difference of the GTGR and IAR is given by
\begin{equation}
\langle E_{\mathrm{GTR}} \rangle -\langle E_{\mathrm{IAR}} \rangle 
=\Delta E_{ls} + 2(\tilde{\kappa}_{\sigma \tau} -\tilde{\kappa}_\tau)\dfrac{N-Z}{A},
\end{equation}
where $\Delta E_{ls}$ is an average value of the spin-orbit 
splitting, and $\tilde{\kappa}_{\sigma \tau}$ and $\tilde{\kappa}_{\tau}$ 
are the coupling constants of the spin-isospin and isospin residual forces 
in the separable Hamiltonian. 
The result shown in Fig.~\ref{fig:GR_energy}(b) suggests that 
our microscopic calculation obeys this simple relation in a good approximation 
as long as the deformed systems are considered. 
The slope parameters $\tilde{\kappa}_{\sigma \tau}-\tilde{\kappa}_\tau$
fitted for SkM* and SLy4 in the deformed Zr isotopes 
are $-16.7$ and $-15.6$ MeV, respectively. 
The slope parameters microscopically obtained here are not 
far from the systematic value of $-14.5$ MeV~\cite{nak82}. 

\begin{figure}
\begin{center}
\includegraphics[scale=0.32]{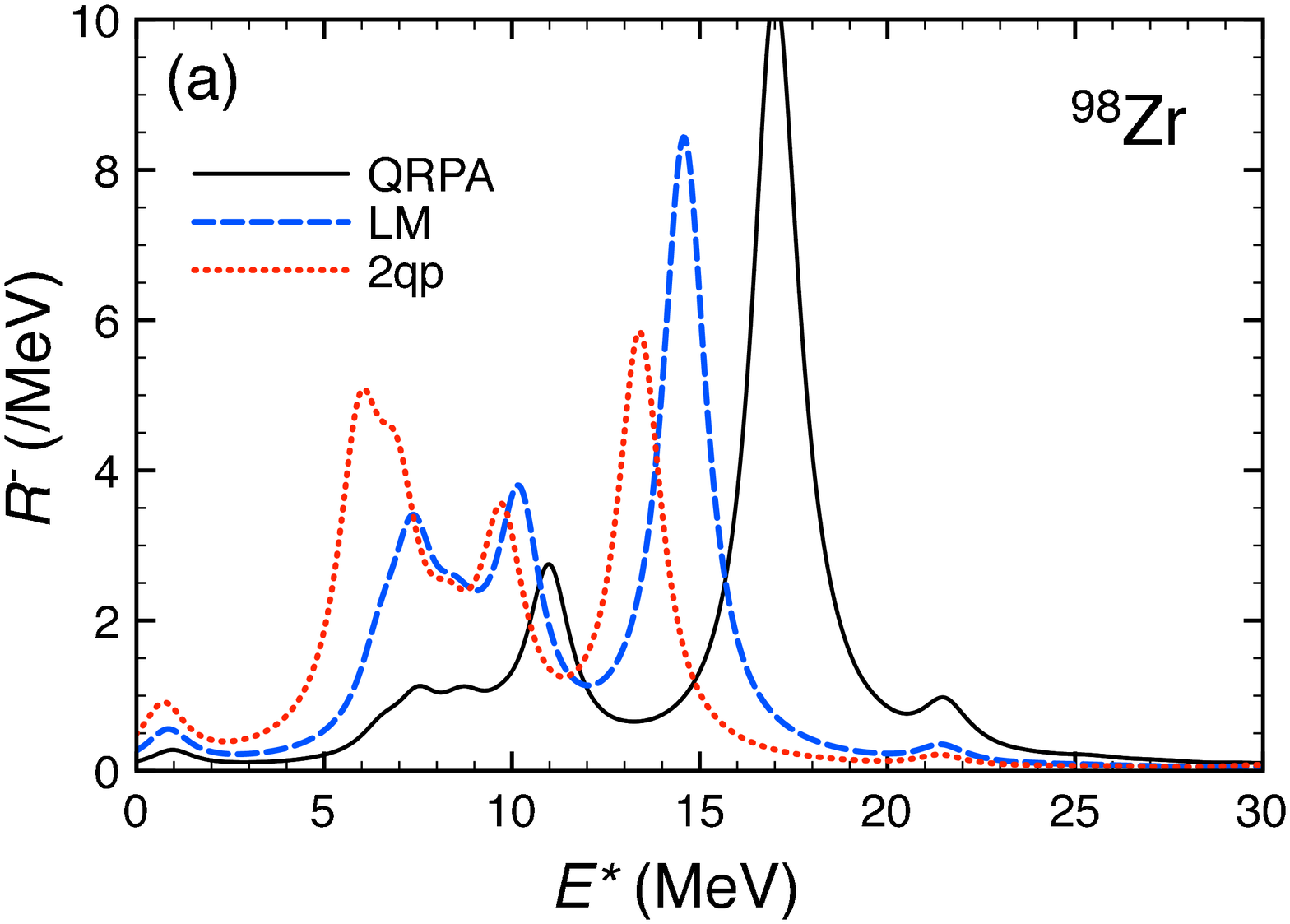}
\includegraphics[scale=0.32]{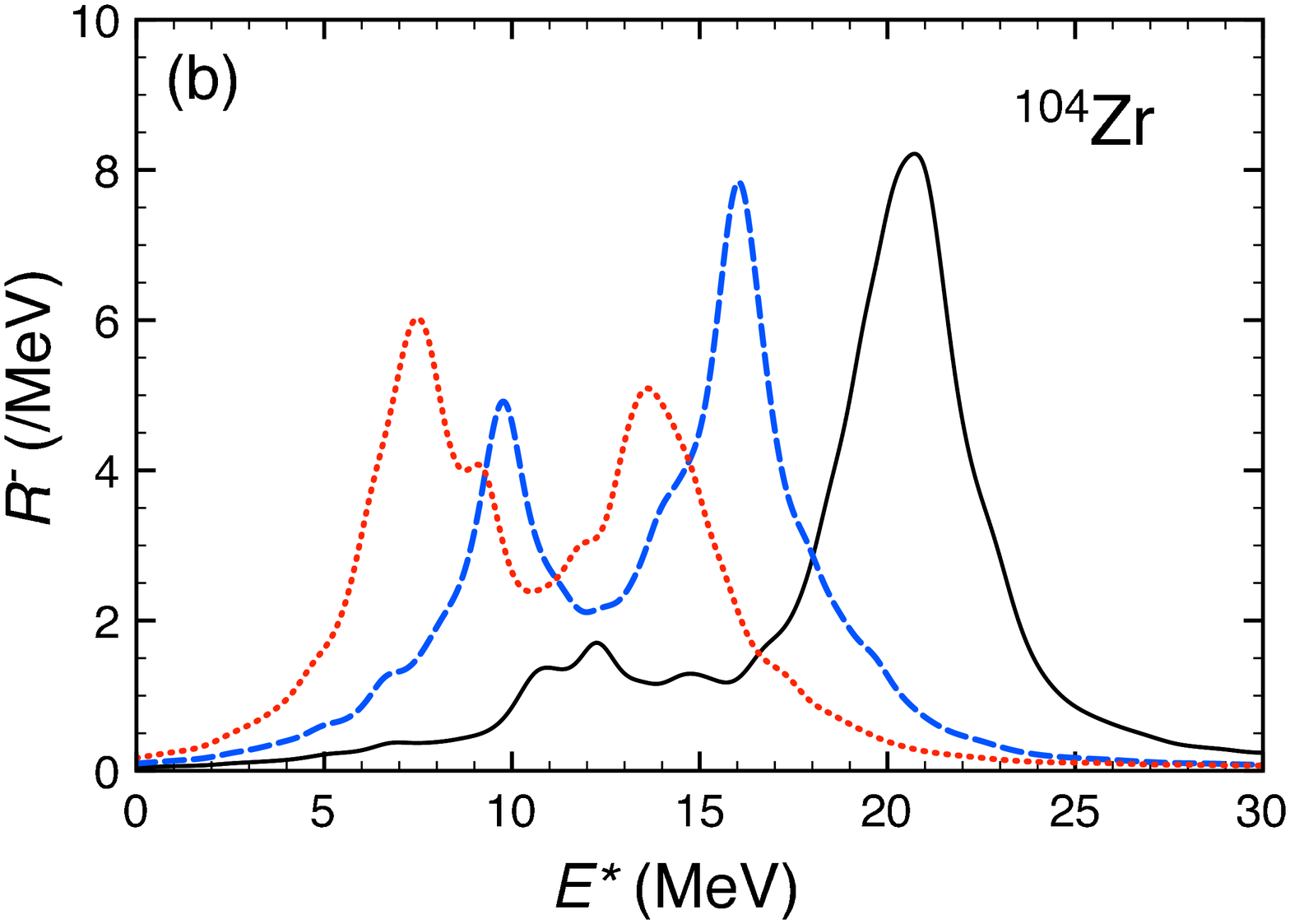}
\caption{
Strength distributions in (a) $^{98}$Zr and in (b) $^{104}$Zr 
calculated by employing the SLy4 EDF with $\gamma$=1 MeV. 
The QRPA strengths are compared 
with the results obtained in the Landau-Migdal approximation (LM) and 
without the residual interactions (2qp). 
\label{fig:104Zr_LM}} 
\end{center}
\end{figure}

Figure~\ref{fig:104Zr_LM} shows the strength distributions 
in some selected isotopes calculated with the SLy4 EDF. 
In the deformed isotopes other than $^{104}$Zr, we see the similar 
features discussed below. 
Here, 
we compare the QRPA results with those obtained 
in the LM approximation, and those obtained 
without the residual interactions.
In the LM approximation, we treat the p-h residual interaction as
\begin{align}
&V_{\mathrm{ph}}(\boldsymbol{r}_1\sigma_1 \tau_1,\boldsymbol{r}_2 \sigma_2 \tau_2) \notag \\ 
&= N_0^{-1}
\left[ 
f_0^\prime \tau_1 \cdot \tau_2
+
g_0^\prime \sigma_1 \cdot \sigma_2 \tau_1 \cdot \tau_2
\right]
\delta(\boldsymbol{r}_1 -\boldsymbol{r}_2)
\end{align}
instead of (\ref{v_res_ph}). 
Here,  $N_0$ is the density of states and the LM parameters $f_0^\prime, g_0^\prime$ 
are deduced from the same Skyrme force which generates the mean field~\cite{gia81}. 
The Fermi momentum $k_F$ appearing in the LM parameters is 
evaluated in the local density approximation.

In both nuclei, one of which is spherical and the other is deformed, 
we see two prominent peaks at around 6 and 14 MeV in the unperturbed 2qp 
transition-strength distribution. 
A difference to be noticed is that 
the strengths are fragmented in $^{104}$Zr due to deformation 
at the mean-field level. 
Associated with the repulsive p-h residual interaction, 
most of the strengths are absorbed by the GTGR, 
and the resonance peak is shifted higher in energy.
The energy shift due to the RPA correlation is much larger in 
$^{104}$Zr than in $^{98}$Zr. 
It is pointed out that the energy and collectivity of the GTGR 
are changed when omitting the momentum-dependent terms in 
the residual interaction for the SLy5 EDF 
in a framework of the spherical HF-BCS + pnQRPA~\cite{fra07}.
We clearly see here that the momentum dependence 
in the p-h residual interaction has a significant effect in 
generating collectivity of the GTGR for the SkM* and SLy4 EDFs.

\subsection{$T=0$ pairing in the GT excitation and $\beta$-decay rate}\label{T=0_pairing}

\begin{figure}
\begin{center}
\includegraphics[scale=0.59]{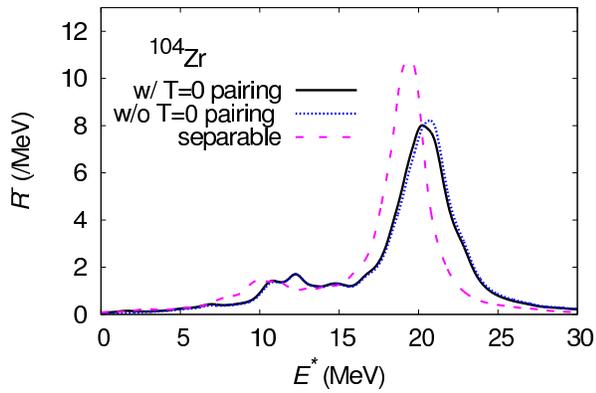}
\caption{
Strength distributions in $^{104}$Zr calculated by employing 
the SLy4 EDF combined with and without the $T=0$ pairing interaction. 
The result in Ref.~\cite{sar10} employing the separable interaction is also shown. 
\label{fig:GT_T=0_104Zr}} 
\end{center}
\end{figure}

\begin{figure}
\begin{center}
\includegraphics[scale=0.49]{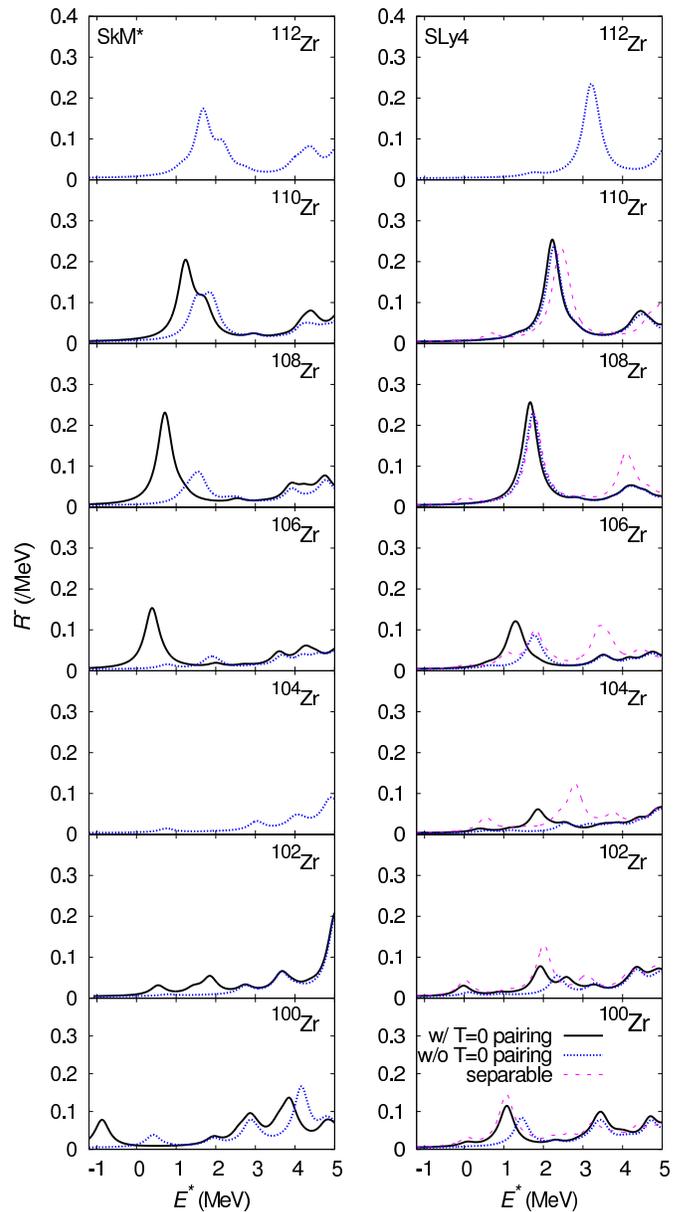}
\caption{
Same as Fig.~\ref{fig:GT_T=0_104Zr} but in the low-excitation energy region 
in the deformed Zr isotopes calculated with the SkM* (left) and SLy4 (right) EDFs 
combined with and without the $T=0$ pairing interaction. 
The results in Ref.~\cite{sar10} are also shown. 
The smearing width $\gamma$ is set $0.1$ MeV.
\label{fig:GT_T=0}} 
\end{center}
\end{figure}

The $T=0$ pairing interaction is effective for the GT 
excitation in the systems where 
the ground states have the $T=1$ pairing condensates. 
Both neutrons and protons are paired in 
$^{104}$Zr with the SLy4 and the mixed-type pairing EDF. 
Thus, we expect to see effects of $T=0$ pairing on the GT strength 
distribution in $^{104}$Zr.

Figure~\ref{fig:GT_T=0_104Zr} shows 
the strength distributions associated with the 
GT$^-$ operator (\ref{eq:GT-}) with the SLy4 EDF 
combined with and without the $T=0$ pairing interaction in $^{104}$Zr. 
We see only a tiny amount of energy change 
due to $T=0$ pairing for the GTGR. 
The change in the centroid energy of the GTGR due to 
$T=0$ pairing is 0.14 MeV. 
This result indicates that the GTGR is built almost entirely of 
the p-h excitations. 

In Fig.~\ref{fig:GT_T=0_104Zr}, we also show the 
result in Ref.~\cite{sar10}, where the 
separable forces were employed for the residual interactions. 
As is discussed in the previous subsection,  
the momentum-dependent terms in the p-h residual interaction 
plays an important role in generating the collectivity.  
The peak energy calculated in Ref.~\cite{sar10} is lower 
than our result by about 1.5 MeV. 

Compared to the GTGR, the low-lying GT strength distribution 
is affected appreciably by the $T=0$ pairing interaction 
due to the following mechanism 
showing up from the structure of the matrix elements 
of the pnQRPA Hamiltonian (\ref{A_matrix_QRPA}) and (\ref{B_matrix_QRPA}). 
The proton orbitals around the Fermi level are partially occupied 
due to the $T=1$ pairing correlations. 
Thus, the neutrons in the hole(-like) orbital can have a chance to 
decay into the proton orbitals through the p-h residual interaction 
and the $T=0$ p-p interaction simultaneously.  
Similarly, when the neutrons are paired, 
the neutrons around the Fermi level 
can decay to protons in the particle(-like) orbitals through 
the p-h and p-p residual interactions. 
We are considering here the $t_-$ channel, but it also holds 
for the $t_+$ channel.

Figure~\ref{fig:GT_T=0} shows the strength distribution 
in the low-excitation energy region in the deformed Zr isotopes. 
In this figure, the smearing width $\gamma$ is set $0.1$ MeV. 
The low-lying states are sensitive to the shell structure around the Fermi levels, 
however, we see some generic features; 
the peak position is shifted lower in energy and at the same time the 
transition strength increases.  

In a well-deformed nuclei, the asymptotic quantum numbers of 
a single-particle orbital are approximately good quantum numbers. 
Though the selection rules based on 
the Nilsson wave functions are broken in a loosely bound system 
as pointed out in Ref.~\cite{yos06}, 
they serve as a zeroth order guideline for 
understanding the structure of the excitation modes. 
For the GT$^-$ operator, the nonvanishing matrix elements 
in $N>Z$ nuclei are given as~\cite{urk01}

\begin{equation}
|\langle \pi[N n_3 \Lambda]\Omega=\Lambda \pm 1/2|
t_- \sigma_{\pm 1}| 
\nu[N n_3 \Lambda]\Omega = \Lambda\mp 1/2 \rangle | = \sqrt{2}.
\label{sele_rule}
\end{equation}

In the strength distribution of $^{100}$Zr calculated with SkM*, 
we see a prominent peak at $E^* \simeq -1$ MeV. 
The QRPA frequency of the lowest $K=1$ state 
is $\omega=0.33$ MeV, 
and the sum of backward-going amplitude squared $\sum Y^2$ is 0.49. 
This eigenstate is predominantly generated by 
a $\nu[422]3/2 \otimes \pi [422]5/2$ excitation. 
Noted that this is a h-h type excitation because 
the occupation probability of 
a $\pi [422]5/2$ orbital is 0.78. 
It indicates that 
this mode has large transition strengths for the proton-neutron-pair 
creation/annihilation operators as well.

In $^{106}$Zr, we see an appreciable effect of the $T=0$ pairing interaction 
on the low-lying state. With SkM*, 
the $K^{\pi}=1^+$ eigenstate at $\omega=1.71$ MeV ($E^*=0.39$ MeV) 
possessing a strength of 0.22 is mainly 
constructed by a $\nu[413]5/2 \otimes \pi[413]7/2$ excitation 
with an amplitude of $X^2-Y^2=0.75$, and 
a $\nu[402]5/2 \otimes \pi [413]7/2$ excitation with an amplitude of 0.14. 
The occupation probability of a $\nu[413]5/2$ orbital is 0.12 
and that of a $\nu[402]5/2$ orbital is 0.01.
Thus, these 2qp excitations are a p-p type excitation, and 
they are sensitive to the residual pairing interaction.

Contrastingly, in $^{110}$Zr, 
the effect of the $T=0$ pairing interaction is small. 
The $K^{\pi}=1^+$ state appearing at $\omega=2.70$ MeV ($E^*=1.23$ MeV) 
is constructed similarly in $^{106}$Zr by 
a $\nu[413]5/2 \otimes \pi[413]7/2$ excitation with a weight of 0.38, 
and a $\nu[402]5/2 \otimes \pi [413]7/2$ excitation with a weight of 0.48. 
In $^{110}$Zr, 
the occupation probability of a $\nu[413]5/2$ orbital is 0.78 and 
and that of a $\nu[402]5/2$ orbital is 0.05. 
Thus, the residual pairing interaction is less effective than in $^{106}$Zr. 
With SLy4, 
the $K^{\pi}=1^+$ state at $\omega=4.06$ MeV ($E^*=2.23$ MeV) 
possessing a strength of 0.39 is also 
mainly constructed by a $\nu [413]5/2 \otimes \pi [413]7/2$ excitation 
with a weight of 0.79, 
and a $\nu[402]5/2 \otimes \pi [413]7/2$ excitation with a weight of 0.17. 
The occupation probability of a $\nu [413]5/2$ orbital is 0.81, 
and that of $\nu[402]5/2$ is 0.16. 
Therefore, 
the the low-lying mode in $^{110}$Zr calculated with SLy4 
is dominantly a p-h type excitation and  
the p-p residual interaction does not play a significant role.

From this analysis, we come to the following conclusion: 
The number of 2qp excitations generating the low-lying mode 
is small in the 
Zr isotopes under consideration. 
The $K^{\pi}=1^+$ state possessing an appreciable strength is 
generated by mainly a 2qp excitation satisfying the 
selection rule (\ref{sele_rule}). 
The effect of the p-p residual interaction in the low-lying mode thus 
depends on the location of the chemical potential, 
whether it is a p-h type or a p-p type excitation.

We also show the results in Ref.~\cite{sar10} in Fig.~\ref{fig:GT_T=0}. 
Although the GTGR are predicted lower in energy in a separable approximation 
than by our calculations, the low-lying strength distributions are not very different. 
In a separable approximation, they have some small strengths 
in the energy region of $0-1$ MeV in all the isotopes.

The low-lying GT strength distribution strongly affects the $\beta$-decay rate. 
Thus, we can clearly see the effect of $T=0$ pairing in the $\beta$-decay life time. 
The $\beta$-decay half life $T_{1/2}$ can be calculated with the 
Fermi Golden rule as~\cite{gov71},
\begin{align}
\dfrac{1}{T_{1/2}} &= \dfrac{\lambda_\beta}{\log 2} \notag \\
&=\dfrac{(g_A/g_V)^2_{\mathrm{eff}} }{D}\sum_{K} \sum_{E_i^* < Q_\beta}f(Z, Q_\beta-E_i^*)
| \langle i | \hat{F}_K^- | 0\rangle  |^2, 
\label{beta_rate}
\end{align}
where $D=6163.4$ s and we set $(g_A/g_V)_{\mathrm{eff}}=1$ rather than its actual 
value of 1.26 to account for the quenching of spin matrix in nuclei~\cite{boh75}. 
The Fermi integral $f(Z,Q_\beta-E_i^*)$ in (\ref{beta_rate}) 
including screening and finite-size effects is given by
\begin{equation}
f(Z,W_0) = \int_1^{W_0} p W(W_0 - W)^2 \lambda (Z,W) dW,
\end{equation}
with
\begin{equation}
\lambda(Z,W)=2(1+\gamma)(2pR)^{-2(1-\gamma)}e^{\pi \nu}
\left|\dfrac{\Gamma(\gamma + i\nu)}{\Gamma(2\gamma +1)}\right|^2,
\end{equation}
where $\gamma=\sqrt{1-(\alpha Z)^2}$, $\nu=\alpha ZW/p$, $\alpha$ is the fine structure 
constant, $R$ is the nuclear radius. 
$W$ is the total energy of $\beta$ particle, $W_0$ is the total energy available in $m_e c^2$ units, 
and $p=\sqrt{W^2-1}$ is the momentum in $m_e c$ units~\cite{gov71}. 
Here, the energy released in the transition from the ground state of the target nucleus 
to an excited state in the daughter nucleus is given approximately by~\cite{eng99}
\begin{equation}
Q_\beta -  E_i^* \simeq \lambda_\nu - \lambda_\pi + \Delta M_{n-H} - \omega_i.
\end{equation}

\begin{figure}
\begin{center}
\includegraphics[scale=0.35]{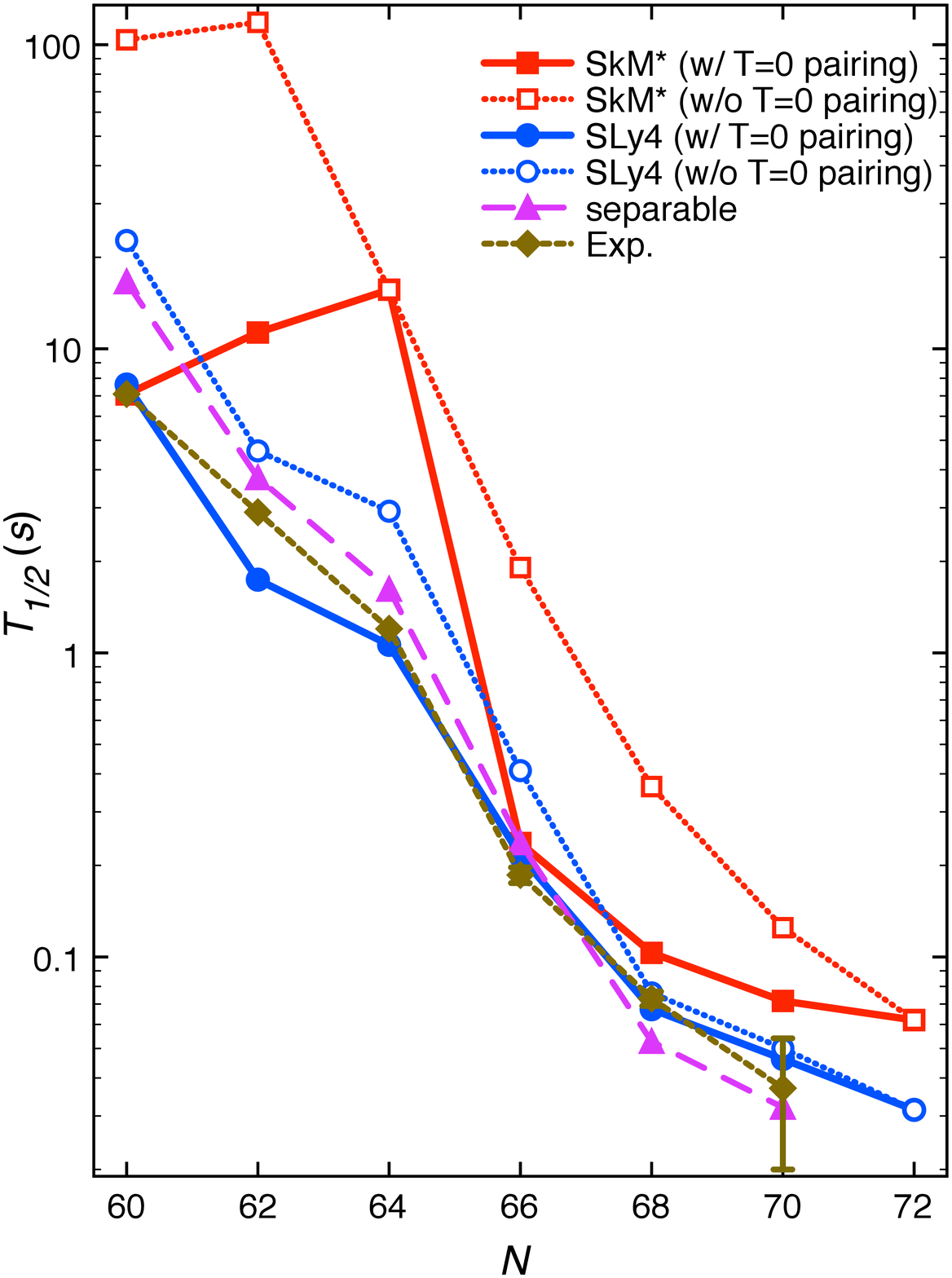}
\caption{
Calculated $\beta$-decay half-lives of the Zr isotopes with the SkM* and SLy4 EDFs 
combined with and without the $T=0$ pairing interaction. 
The results in Ref.~\cite{sar10} and experimental data~\cite{Zr100,Zr102,Zr104,nis11} are 
also shown. 
\label{fig:halflives}} 
\end{center}
\end{figure}

Figure~\ref{fig:halflives} shows the the $\beta$-decay half lives of 
the Zr isotopes thus calculated with the SkM* and SLy4 EDFs combined 
with and without the $T=0$ pairing interaction. 
We see that the attractive $T=0$ pairing interaction shortens substantially 
the $\beta$-decay half lives. 

Without the $T=0$ pairing interaction, 
the half lives calculated with SkM* 
are about $2-20$ times longer than those with SLy4. 
About a half of the differences is 
due to the smaller $Q_\beta$ values calculated with SkM* than with SLy4. 
Thus, we need a stronger p-p interaction for SkM* 
to reproduce the observed half life of $^{100}$Zr.
Then, the half lives calculated with the two EDFs together with the $T=0$ pairing 
interaction come closer to each other exept in $^{102,104}$Zr. 
As mentioned in Sec.~\ref{result_gs}, 
each of the isotopes has the different pairing properties depending 
on the Skyrme-EDF employed.

In Fig.~\ref{fig:halflives}, we include the results in Ref.~\cite{sar10} 
together with the available experimental data~\cite{Zr100,Zr102,Zr104,nis11}. 
The recent experiment data obtained at RIKEN RIBF 
show the short half lives in the $^{110}$Zr region~\cite{nis11}. 
They are reproduced well by the calculation in Ref.~\cite{sar10}, 
while it overestimates the half lives of 
the lighter Zr isotopes. 
A good reproduction of the half lives in the $A \sim110$ region 
may be due to the presence of the low-lying strengths.  
Our calculations, in particular employing SLy4, 
reproduce well the observed half lives systematically in $^{100-110}$Zr.

The low-lying GT strengths relevant to the $\beta$-decay rate 
are a quite delicate quantity because they emerge 
as a consequence of 
cancellation between a repulsive p-h residual interaction 
and an attractive $T=0$ pairing interaction, both of which are largely uncertain 
in a nuclear EDF method. 
We need a more reliable EDF together with the proton-neutron pairing interaction 
that is able to describe well the spin-isospin excitations systematically 
in order to make further steps toward 
a self-consistent and systematic description of 
$\beta$-decay of nuclei involved in the $r$-process nucleosynthesis. 
A virtue of 
our new framework developed in the present article 
is that it is suitable for a systematic calculation of the 
spin-isospin responses of nuclei 
because it is applicable to a nucleus with an arbitrary mass number
whichever it is spherical or deformed, deeply bound or weakly bound 
in a reasonable calculation time with a help of the massively parallel computers, 
once the EDF and the proton-neutron pairing interaction are given. 
Through the systematic calculations employing several parameter sets 
of interaction 
and their comparison with available 
experimental data or observations, 
we can put constraints on the 
spin-isospin part of the new EDF and the proton-neutron 
pairing interaction. 
Then, we can proceed to a more reliable calculation.

\section{Summary}\label{summary}

We have developed the fully-self-consistent 
framework to calculate the spin-isospin collective modes 
of excitation in nuclei using the Skyrme EDF.
We solve the deformed HFB equations on a grid in coordinate space. 
This enables us to investigate the excitation modes 
in nuclei off the stability line with an arbitrary mass.

Numerical applications have been performed for 
the Gamow-Teller excitation in the deformed neutron-rich Zr isotopes. 
We found a small amount of fragmentation due to deformation 
in the GT transition-strength distribution.
The momentum-dependent terms in 
the p-h residual interaction play an important role 
in generating the collectivity. 
An attractive $T=0$ pairing interaction 
has little influence on the energy 
of the GT giant resonance, while lowers the energies and 
enhances the GT strengths in the low energy region.
The effect of the $T=0$ pairing interaction 
in the low-lying mode depends sensitively on 
the location of the Fermi level of neutrons.

$\beta$-decay rates depend primarily on the $Q_\beta$ value, 
the residual interactions 
for both the p-h and the p-p channels, and the shell structures. 
The framework developed in this article treats 
self-consistently these key ingredients on the same footing. 
Once the strength of the $T=0$ pairing interaction is determined 
so as to reproduce the observed $\beta$-decay half-life of $^{100}$Zr, 
our calculation scheme produces well the isotopic dependence 
of the half lives up to $^{110}$Zr as was recently observed at RIKEN RIBF.

Systematic calculations with the HFB + pnQRPA
for nuclei in a whole nuclear chart 
help us not only to understand and to predict new types of collective modes of
excitation in unstable nuclei, 
and to provide the microscopic inputs for the astrophysical simulation 
but also to shed light on the nuclear EDF of new generations.

\begin{acknowledgments}  
The author thanks P.~Sarriguren for providing him with 
the GT strength distributions to be compared. 
He also thanks K.~Matsuyanagi, N.~Van Giai, 
H.~Z.~Liang and F.~Minato for stimulating discussions.
This work was supported by KAKENHI Grant Nos. 23740223 and 25287065. 
The numerical calculations were performed 
on SR16000
at the Yukawa Institute for Theoretical Physics, Kyoto University and 
on T2K-Tsukuba, at the Center for Computational Sciences, University of Tsukuba. 
Part of the results is obtained by using the K computer at the RIKEN
Advanced Institute for Computational Science and by pursuing HPCI Systems
Research Projects (Proposal Number hp120192).

\end{acknowledgments}

\end{document}